# Mini-step Strategy for Transient Analysis

Fei Wei, Huazhong Yang

Electronic Engineering Department, Tsinghua University, Beijing, China, 100084

2011-3-12

## Abstract

Domain decomposition methods are widely used to solve sparse linear systems from scientific problems, but they are not suited to solve sparse linear systems extracted from integrated circuits. The reason is that the sparse linear system of integrated circuits may be non-diagonal-dominant, and domain decomposition method might be unconvergent for these non-diagonal-dominant matrices. In this paper, we propose a mini-step strategy to do the circuit transient analysis. Different from the traditional large-step approach, this strategy is able to generate diagonal-dominant sparse linear systems. As a result, preconditioned domain decomposition methods can be used to simulate the large integrated circuits on the supercomputers and clouds.

## 1. Introduction

An integrated circuit can be modeled by a nonlinear ordinary differential equation (1) [1]. $x$ is the node's voltage, $C$ is the capacitor matrix, $i_s$ is the current source. $f(x,t)$ is the nonlinear conductor matrix.

$$C\frac{dx}{dt} + f(x,t) = i_s \tag{1}$$

Then we use backward Euler method to discretize equation (1) into a nonlinear equation (2).

$$C\frac{x^k - x^{k-1}}{\Delta t} + f(x^k) = i_s \tag{2}$$

Here $\Delta t$ is the time step.

Later, we use Newton-Raphson method to solve equation (2). First we estimate $f(x^k)$ by Taylor expansion in equation (3).

$$f(x^k) = f(x^{k-1}) + f'(x^{k-1})(x^k - x^{k-1}) \tag{3}$$

After that, we insert equation (3) into equation (2):

$$\frac{C}{\Delta t}(x^k - x^{k-1}) + f(x^{k-1}) + f'(x^{k-1})(x^k - x^{k-1}) = i_s \tag{4}$$

$$\left(\frac{C}{\Delta t} + f'(x^{k-1})\right)(x^k - x^{k-1}) = i_s - f(x^{k-1}) \tag{5}$$

Finally, we make some simplification to (5), and obtain (6):

$$\left(\frac{C}{\Delta t}+G\right)\Delta x = i_s - F \qquad (6)$$

Here $\Delta x = x^k - x^{k-1}$. $G = f'(x^{k-1})$, and $F = f(x^{k-1})$.

According to (6), C is a diagonal dominant matrix, and G is an unsymmetric non-diagonal dominant matrix. If $\Delta t$ is smaller enough, then $\frac{C}{\Delta t}+G$ can be a diagonal dominant matrix. This is the idea of mini-step strategy.

## 2. Maxmimum Step Estimation

In this section, we calculate the maximum time step $\Delta t$ for a certain semiconductor technology. MOS1 model is used to construct equation (6).

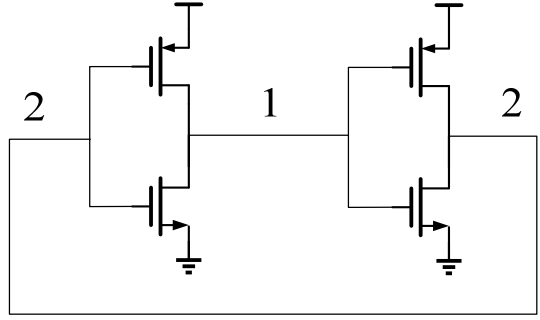

Figure 1. Two inverters connected end-to-end

Assume that there are two inverters connected end-to-end, as shown in Fig. 1. We mainly focus on the worst case, when the off-diagonal elements in $G$ are the largest.

$$C = \begin{bmatrix} 2C_g & \\ & 2C_g \end{bmatrix}$$

$$G = \begin{bmatrix} 0 & -2g_m \\ -2g_m & 0 \end{bmatrix}$$

Here $C_g = W*L*C_{ox}$, $g_m = K_p \frac{W}{L}(V_{gs} - V_{TH})$. $W$ and $L$ is the width and length of the transistor, respectively. $C_{ox}$ is the gate oxide capacitor of unit area. $K_p$ is the trans-conductance parameter. $V_{TH}$ is the threshold voltage. $V_{gs}$ is the voltage between the gate and source of the transistor.

To make $\frac{C}{\Delta t}+G$ diagonal-dominant,

$$\frac{2C_g}{\Delta t} > 2g_m$$

Then we get:

$$\frac{2W*L*C_{ox}}{\Delta t} > 2K_p \frac{W}{L}(V_{gs} - V_{TH})$$

$$\Delta t < \frac{C_{ox}}{K_p} L^2 (V_{gs} - V_{TH})$$

Here we assume that $V_{gs} - V_{TH} = 0.5*V_{dd}$, $V_{dd}$ is the supply voltage, so,

$$\Delta t < \frac{C_{ox}}{2K_p} L^2 \cdot V_{dd} \qquad (8)$$

Because the carrier mobility $\mu_0 = \frac{K_p}{C_{ox}}$, then,

$$\Delta t < \frac{L^2}{2\mu_0} V_{dd} \qquad (9)$$

Thus we know that, the maximum time step is dependent on the minimum length of the MOS transistor. For the 100nm technology, $L = 100n$ m, $\mu_0 = 1.0$ m$^2$/(V s), $V_{dd} = 1.0$ V, as a result, $\Delta t < 5f$ s. This is a pessimistic estimation of the maximum step.

## 3. Conclusion

This paper proposes a mini-step strategy for parallel circuit simulation. The benefit of this approach is that the sparse matrix of circuits will be diagonal dominant. Diagonal dominant matrices are easy to be solved by iterative algorithms [2]. The penalty of this approach is that smaller steps will lead to much more computations, which will slow down the simulation significantly.

This mini-step strategy is compatible with the domain decomposition method proposed in [3].

## References

1. L. Nagel. SPICE2: A Computer Program to Simulate Semiconductor Circuits, Electronics Research Laboratory Report No. ERL-M520. University of California, Berkeley, 1975.
2. He Peng, Chung-Kuan Cheng. Parallel transistor level full-chip circuit simulation. DATE 2009, 304-307.
3. Fei Wei, Huazhong Yang. Transmission Line Inspires a New Distributed Algorithm to Solve the Nonlinear Dynamical System of Physical Circuits. arXiv.org, 2010.